\documentclass{npw}
\usepackage{graphicx}
\usepackage{color}	

\begin{document}

\title{The Negative Parity Bands in \boldmath $^{156}$Gd}
\author{M Jentschel\email{jentsch@ill.fr} \\
  \it Institut Laue-Langevin, BP 156, 6 Rue Jules Horowitz, 38042 Grenoble Cedex 9, France \\[3mm]
  L Sengele, D Curien, J Dudek and F Haas \\
  \it Institut Pluridisciplinaire Hubert Curien, 3 Rue du Loess, BP28, 67037 Strasbourg Cedex 2, France }
\pacs{21.10.Dr, 21.10.Ma, 21.60.Cs, 21.60.Jz, 25.85.Ca}
\date{}
\maketitle

\begin{abstract}
The high flux reactor of the Institut Laue-Langevin is the world most intense neutron source for research.  Using the ultra high-resolution crystal spectrometers GAMS installed at the in-pile target position H6/H7 it is possible to measure nuclear state lifetimes using the Gamma Ray Induced Recoil (GRID) technique. In bent crystal mode, the spectrometers allow to perform spectroscopy with a dynamic range of up to six orders magnitude. At a very well collimated external neutron beam it is possible to install a highly efficient germanium detector array to obtain coincidences and angular correlations.
The mentioned techniques were used to study the first two negative parity bands in $^{156}$Gd. These bands have been in the focus of interest since they seem to show signatures of a tetrahedral symmetry. A surprisingly high B(E2) value of about 1000 W.u. for the $4^- \rightarrow 2^-$ transition was discovered. It indicates that the two first negative parity bands cannot be considered to be signature partners. 
\end{abstract}

\section{Introduction}

Theoretical prediction of the presence of the tetrahedral point group symmetry in nuclei have motivated a series of experimental investigations. The symmetry is expected to open new gaps between the nucleonic orbitals helping to stabilise its shape at a non-zero value of the $\alpha_{32}$ tetrahedral deformation and leading to new  so-called tetrahedral magic gaps: 16, 20, 32, 40, 56, 64, 70, 90, 112 and 136, Refs.~\cite{XLi94,JDu02}. The tetrahedral energy gaps at these nucleon numbers, with the sizes sometimes comparable to those at the spherical magic numbers, correspond to pure tetrahedral deformation ($\alpha_{32}$ in terms of the nuclear surface representation with the help of the spherical harmonic basis). Pure tetrahedral-symmetry shapes generate neither quadrupole nor dipole moments, whereas the fact of being non-spherical generates the rotational bands with the energy-spin dependence as $E_I\sim I(I+1)$. Therefore both, the population and detection, of such rotational states by transitions other than the octupole ones should be considered as very rare events. In other words, the transitions other than E3, can be envisaged for instance as the result of various types of polarisation in terms of shapes either by the valence particles, or by zero-point motion and/or Coriolis effects, but are not expected to be strong. 

According to theoretical predictions, in several nuclei the tetrahedral symmetry minima lie low in the energy scale and compete with the axial quadrupole-deformation ground-state minima. Using the two-dimensional projections of the calculated potential energy  onto the $(\alpha_{20}, \alpha_{32})$ deformation plane one obtains overall relatively flat landscapes. Under these conditions calculations predict large amplitude fluctuations in the $\alpha_{32}$ (tetrahedral) direction around the quadrupole equilibrium as well as the accompanying low vibration energies in the corresponding mode.
Calculations by various theory groups suggest that when the two octupole modes, i.e.~$Y_{32}$ (tetrahedral) and $Y_{30}$ (axial-octupole) come into competition - the tetrahedral mode wins energetically in majority of the studied cases, cf. \cite{JDu13} and references therein.

The actual crucial question is whether this non axial deformation $\alpha_{32}$ is experimentally distinguishable from the axial octupole $\alpha_{30}$ deformation? A particular focus has been on the first negative parity bands of the isotope $^{156}$Gd. First theoretical works \cite{Dud06,God10} suggested that missing E2 transitions between the low spin members of these bands could be considered as an indicator of missing quadrupole deformation and favouring therefore a pure tetrahedral -- in contrast to the tetrahedral-oscillation -- interpretation. A recent measurement employing the Bragg spectroscopy \cite{jen2010}  demonstrated however the presence of weak E2 transitions (see Fig.\,1) still carrying a relatively large quadrupole moment. 

At first this could have been seen as disfavouring the tetrahedral symmetry interpretation generally. 
%for this particular band. 
However, later theoretical work \cite{dob11} showed that the presence of the tetrahedral symmetry is compatible with the presence of some non-zero quadrupole moment if the so-called zero-point motion is taken into consideration.  Moreover -- the tetrahedral component in the nuclear mean field can very well give rise to the `tetrahedral oscillations' of the nuclear ground-states leading to the $K^\pi=2^-$ bands in full analogy to the $K^\pi=2^+$ bands (the well known 
$\gamma$-bands) with the strong quadrupole moments present in both cases. 

The progress just mentioned  has made the simple fingerprint of vanishing E2 strength in negative parity bands obsolete. Therefore the experimental activity has moved to a more systematic investigation of the E1/E2 branching ratios with a particular focus on the so-called signature(simplex)-partner bands. In fact, in $^{156}$Gd the first negative parity band with odd spins has a signature partner with even spins, which -- if the negative-parity band can be associated with the octupole deformation -- should show similar to the odd spin band E1/E2 branching ratios. In this context a series of lifetime, and branching ratio measurements was carried out to investigate the nature of the lowest lying negative parity bands in $^{156}$Gd.

\section{Experimental Setup}
\begin{figure}
\includegraphics[width=\columnwidth]{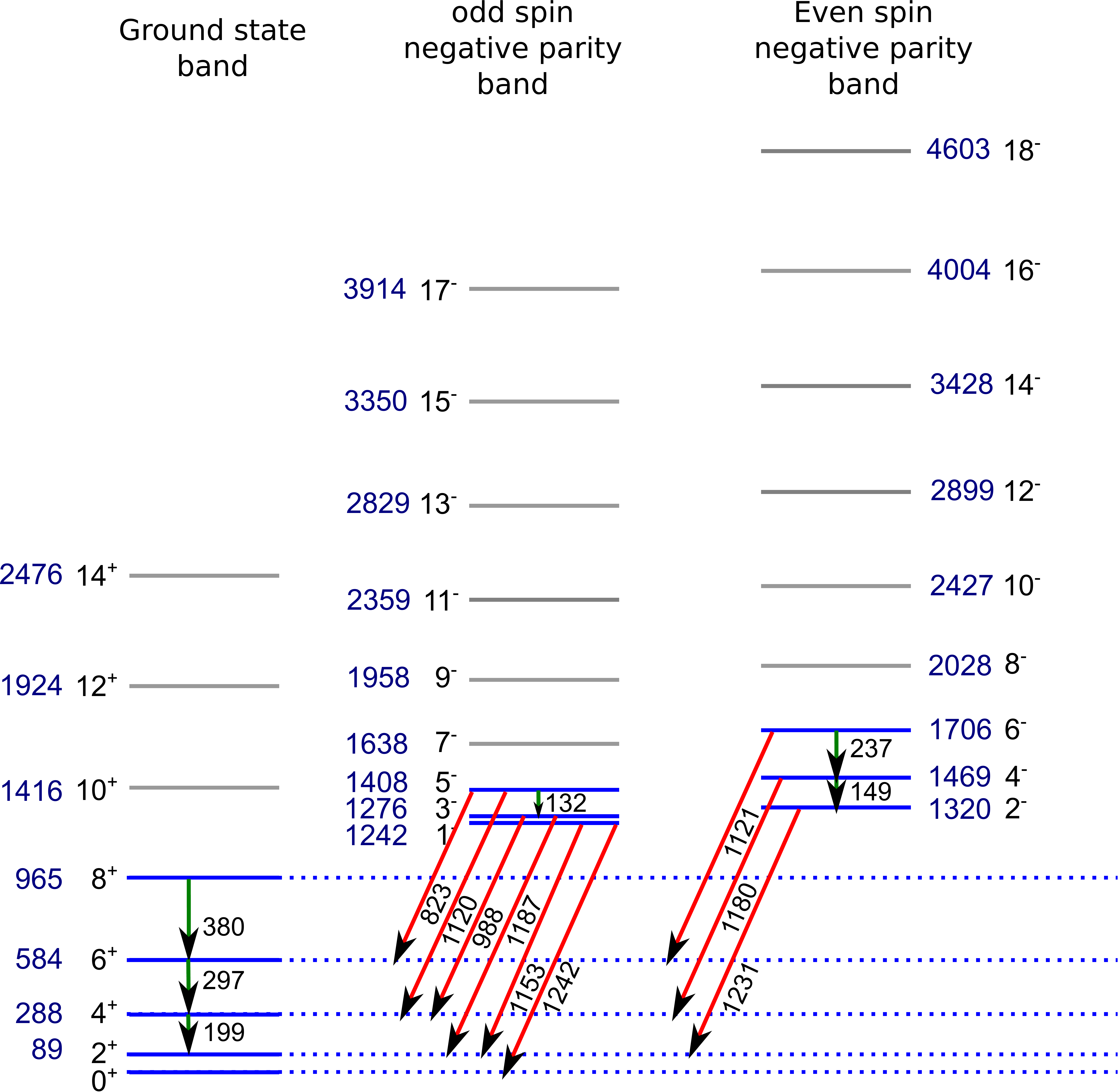}
\caption{\label{f1} The first two negative parity bands in $^{156}$Gd. The levels excited by the $^{155}$Gd(n,$\gamma$)$^{156}$Gd are shown in blue together with E2 transitions in green and E1 transitions in red. Energies, spins and parities are taken from \cite{ensdf}. }
\end{figure}

\begin{figure}
\includegraphics[width=\columnwidth]{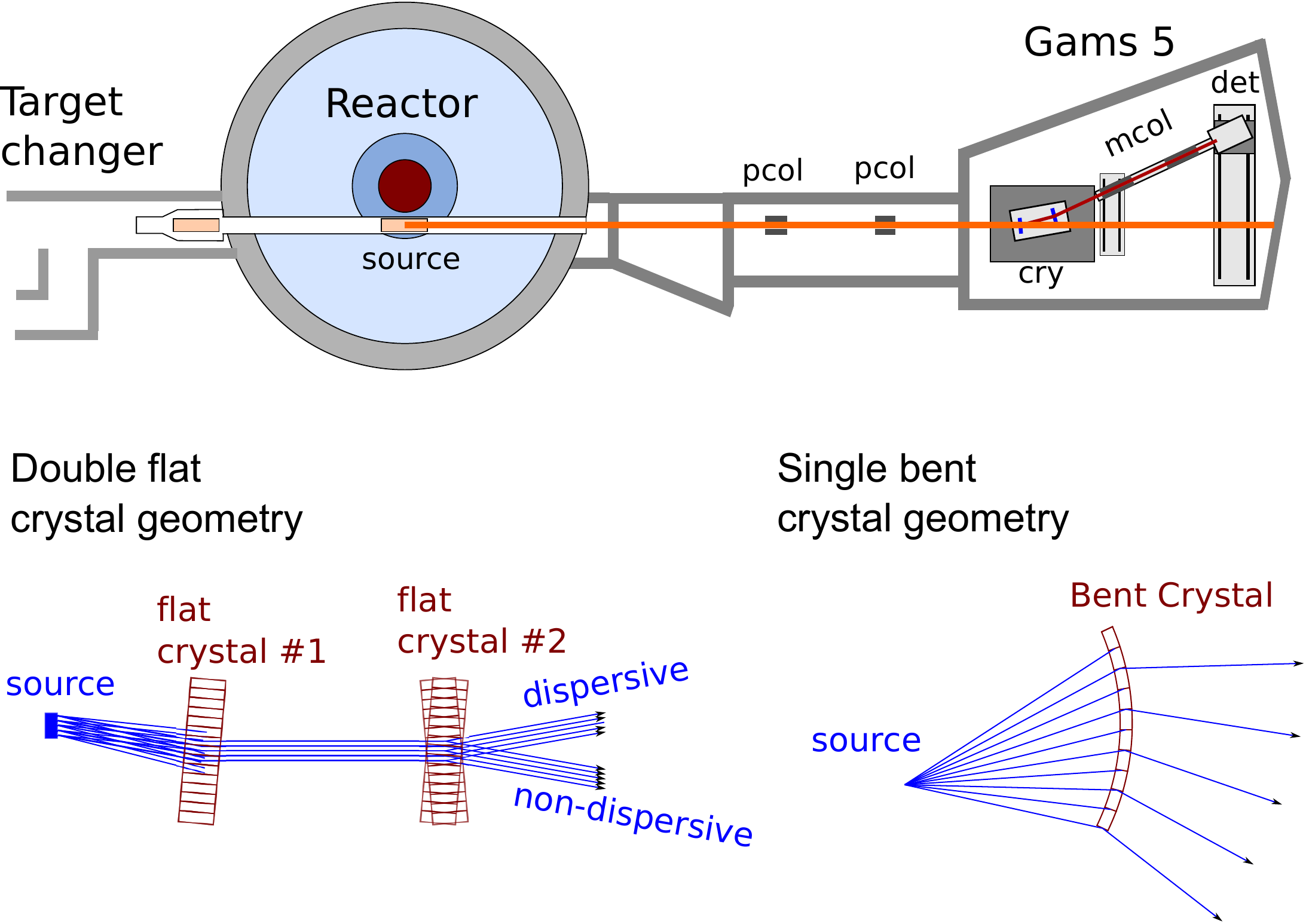}
\caption{\label{f2} The upper part shows a schematic layout of the crystal spectrometer GAMS5, which is placed 17 meters from the in-pile source position. The beam from the source is first pre-collimated (pcol) by a fixed collimation system, than a monochromatized beam is produced by the crystals (cry), separated by a movable collimation system (mcol) from the direct beam and than counted by a detector (det). The indicated diffraction angles are strongly exaggerated for visualisation purposes. In the lower part of the figure, the two crystal diffraction modes and their different acceptance with respect to beam divergence are schematically visualised.  }
\end{figure}

A first experimental investigation of the negative parity bands in $^{156}$Gd with respect to an experimental search of tetrahedral symmetry was carried out by Doan et al. \cite{doa2009} using a fusion-evaporation reaction $^{154}$Sm($\alpha$,2n). The reaction allowed to populate the high spin states of the negative parity bands (up to spin 17 $\hbar$) and the use of the JUROGAM $\gamma$-ray detector array and evaluation of $\gamma\gamma\gamma$ coincidences allowed a clear assignment of all transitions. In the experiment all inter-band E1 transitions from the first two negative parity bands were assigned. Vanishing E2 transitions at the bottom of the odd-spin band were not detected below spin $9^-$ and also the experiment was not able to establish the $4^− \rightarrow 2^−$ transition in the even-spin band. 
Complementary to the experiment of Doan et al., a series of experiments \cite{jen2010,klora} based on the reaction $^{155}$Gd(n,$\gamma$)$^{156}$Gd was carried out. This reaction has a very strong cross section (64000 barn) and populates mostly the lower spin states (below spin 7$\hbar$) of the negative parity bands. Experiments were carried out with the crystal spectrometer GAMS5 in double flat crystal mode (see figure  \ref{f2} and \cite{jen2010}) for the measurement of nuclear state life times, in single bent crystal mode for the measurement of intensities of weak transitions. The reaction was also studied within the EXILL campaign \cite{mutti13}, where a highly efficient HPGe-detector array was placed around a neutron beam. A simplified level scheme - showing the transitions investigated within this work is shown in Figure \ref{f1}.

\subsection{The Crystal Spectrometer GAMS5}
The instrument and its options as double flat and as single bent crystal spectrometer was already described in the context of former publications \cite{jen2010} and therefore only a short summary shall be given here. In the double crystal mode the spectrometer is capable of achieving a relative energy resolution of $\Delta E/E \simeq 10^{-6}$. This extraordinary resolution is achieved for the price of a very small effective solid angle of 10$^{-11}$. This allows  to carry out experiments with massive samples  of several grams of mass only. These samples are introduced into the in-pile beam tube H6/H7 of the research reactor of the Institut Laue-Langevin, where they are exposed to a neutron flux of $5 \times 10^{14}$ neutrons per second and cm$^2$. In double flat crystal mode the instrument can be operated in two diffraction geometries: i) The so called non-dispersive geometry, having the two crystals in a parallel alignment with respect to each other, allows measuring the instrument response function. The measured response function is compared to a theoretical calculation and the deviation is deduced as an universal parameter (essentially determined by the alignment and the vibration amplitude of the crystals).  ii) The dispersive geometry, having between the two crystals a well defined Bragg angle, allows measuring additional -- with respect to the instrument response function -- broadening of the $\gamma$-ray line. The primary source of broadening of a $\gamma$-ray line is Doppler broadening due to the motion of the emitting nuclei. The resolution of the spectrometer is sufficiently good to detect Doppler broadening  associated with  the thermal motion of atoms and this so called thermal Doppler broadening corresponds to the minimum broadening, which can be obtained in a measurement. Another source of Doppler broadening can be observed in the case of a $\gamma$-ray cascade: Every $\gamma$-ray emission is inducing a recoil to the emitting nucleus, which induces a recoil motion of the nucleus being slowed down by inter-atomic collisions. The energy of subsequently, within a sufficiently small time window after the recoil, emitted $\gamma$-rays are Doppler shifted if measured in a laboratory frame. Since the recoil process is isotropic one observes in a measurement a Doppler broadening. The measurement of Doppler broadening of these secondary $\gamma$-rays allows to 
determine the time between $\gamma$-ray emissions - the nuclear lifetime of the intermediate level. This approach to measure nuclear state lifetimes is called the GRID (Gamma Ray Induced Doppler broadening) lifetime technique and is in detail described in a number of publications, \cite{born93, jen00}. Since it essentially requires the best possible energy resolution it can be only realised in double flat crystal mode. Due to the low luminosity of the spectrometer in this mode, massive samples of about 10 grams Gd$_2$O$_3$ powder with natural isotopic abundance were used.

The spectrometer can also be equipped with curved crystals, which 
%helps
help to increase the solid angle by four orders of magnitude. This allowed to use samples of a few tens of milligrams mass of isotopically 95$\%$ enriched samples of $^{155}$Gd$_2$O$_3$. In this configuration the spectrometer has an energy resolution of $\Delta E/E \simeq 10^{-6} \times E {\mbox{[keV]}}$. This means that up to an energy of about 1.5 MeV the resolution is better compared to normal HPGe-detectors. The main advantage of this geometry is, however, the possibility to obtain a very good measurement of relative intensities. This comes essentially from the fact that the detector, used to count the diffracted $\gamma$-rays, is loaded per time only with selected (by the diffraction process) energies. This yields a dynamic range of up to $10^6$, allowing to search for very weak transitions. A former generation of this spectrometer was used to carry out a rather complete spectroscopy of $^{156}$Gd \cite{klora}. Due to the high neutron capture cross section and the high flux at the sample position small sample masses are `burning out' in the reactor within a few days. Since the focus in \cite{klora} was on a complete scan, the measurement was repeated to assure that the intensities of all branching depopulating the negative parity bands were correctly assigned.

Since the solid angle in both diffraction modes is very small it is impossible to consider a crystal spectrometer for coincidence measurements. Therefore a direct assignment of $\gamma$-rays to a particular band has to result from additional measurements with HPGe-detector arrays.

\subsection{The EXILL setup}
Complementary to GAMS5 in bent crystal mode, the use of a HPGe-array offers the higher resolution power for high energies and, most importantly, the possibility to carry out coincidence measurements. The latter option is also quite important for the correct extraction of nuclear state lifetimes via the GRID technique. Since the Doppler broadening is used to extract lifetimes, an important  parameter is the knowledge of the  recoil velocity distribution. It results directly from the knowledge of the feeding of a particular level of interest (LOI). In the majority of cases the published  information about the feeding is rather incomplete. By gating on $\gamma$-rays from below the LOI it is possible to re-construct a large part of the feeding. 

The concept of the EXILL campaign was to install a highly efficient HPGe-detector array around a neutron beam. The ILL research reactor is offering a large number of neutron guide systems allowing to transport neutrons over hundreds of meters to experimental areas. The most intense of these guides is the ballistic super mirror guide H113 with its end position Pf1b. A detailed description of the beam characteristics can be found in \cite{abele06}. The beam guide delivers a thermal neutron capture equivalent flux density of $2 \times 10^{10} \mbox{n s}^{-1}\mbox{cm}^{-2}$ and an angular divergence of about 7 mrad on an exit window of $20 \times 6 \mbox{cm}^2$. The beam profile and its divergence is too large to be directly used in context with a HPGe-array. Therefore a dedicated collimation system was developed allowing to shape the beam five meters downstream from the end of the H113 guide to a circular cross section of 1 cm diameter and a neutron flux of about $1 \times 10^8 \mbox{n s}^{-1}\mbox{cm}^2$. Connected to this collimation system was a target chamber of about 1 meter length crossing the centre of a detector array and followed by a neutron beam dump. Both, collimation  and target chamber were made out of Aluminium and all neutron optical components were made out of B$_4$C or isotopically enriched $^7$LiF.  The target chamber was surrounded by HPGe detector array consisting of 10 EXOGAM clovers, 6 GASP coaxial  and 2 clover detectors from ILL. The entire system was connected to a trigger free digital acquisition system allowing to record all detected events on a common time base and to set coincidence conditions later. The sample material in this experiment was the same as used in the bent crystal mode. However, due to the very high neutron capture cross section and the high efficiency of the detector array the sample mass was reduced to a few powder grains of an immeasurable small mass (below 1 mg). In this configuration the signals from the neutron capture reaction on $^{155}$Gd were still dominating and driving the detectors close to the measurable saturation threshold (about 18 kHz count rate on each channel).

A more detailed description of the EXILL setup can be found in \cite{mutti13}.

\section{Experimental Results}

\begin{figure}
\includegraphics[width=\columnwidth]{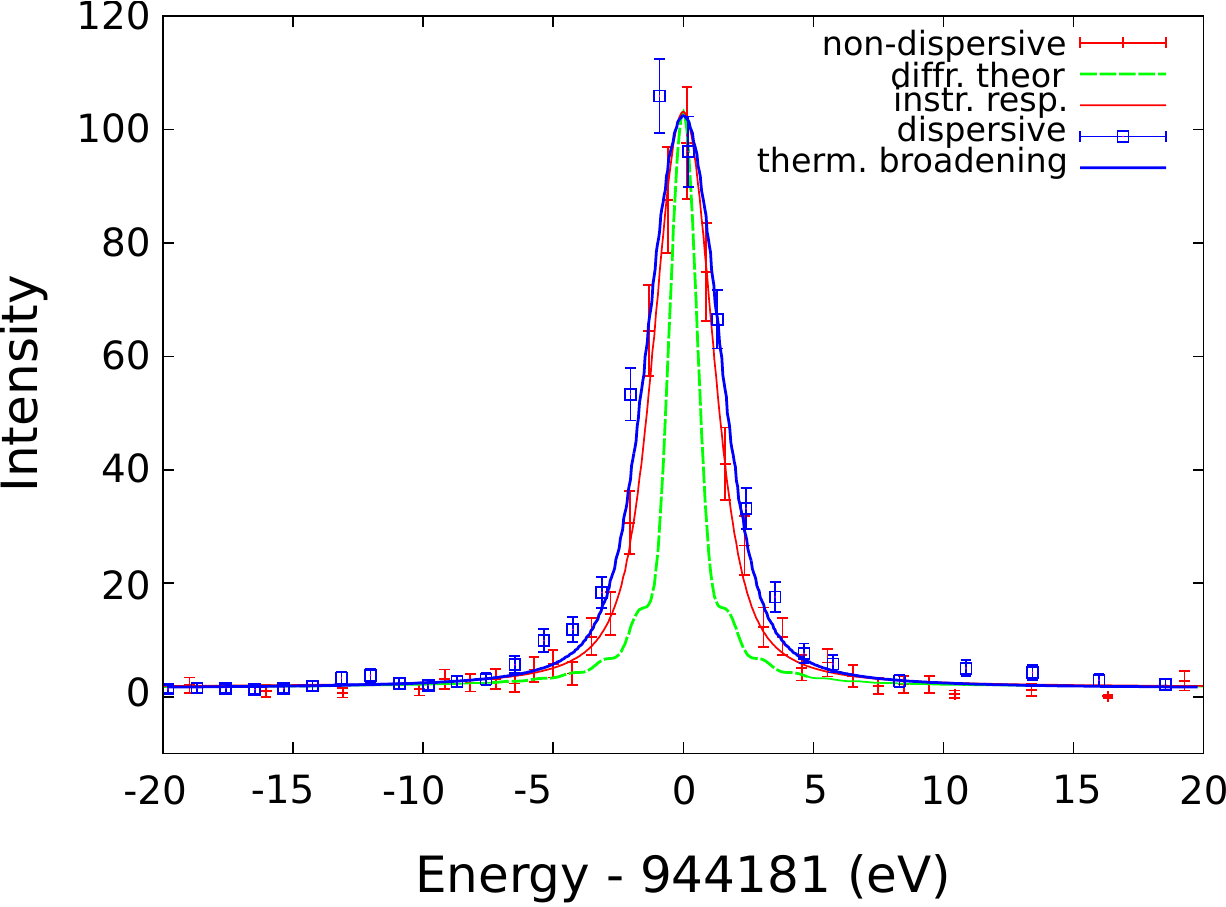}
\caption{\label{f3} Plot of the combined measurements of instrument response function and thermal broadening. The resolution was about 4 eV FWHM at 944.181 keV, essentially dominated by the theoretical diffraction profile. The thermal Doppler broadening is rather low, the average thermal velocity was determined to be 400 m/s.}
\end{figure}

In the earlier work \cite{jen2010}, the odd-spin  negative-parity band was already investigated. The lifetime of the $5^-$ state was measured with the GRID technique to be ${\tau=0.22 \left(^{0.18} _{0.03} \right)\mbox{ps}}$  yielding a $\mbox{B(E2,} 5^- \rightarrow 3^-) = 293\left(^{61} _{134} \right) \mbox{W.u.}$ and a quadrupole moment for this band head to be ${Q_0=7.1 \left(^{0.7}_{1.6} \right)\mbox{b}}$.  The main focus of this work was on the even-spin  negative parity band interpreted by some authors \cite{kon81, cot96, sug11} as the signature partner band of the one with the odd spins. The nuclear state lifetime of the $4^-$ state was measured via the Doppler broadening of the 1180 keV transition and the lifetime of the $2^-$ state via the {1231 keV} transition respectively. A measurement of $6^-$ state was not possible since this state is too weekly populated in the neutron capture reaction and the low solid angle of the double flat crystal mode was not allowing to obtain sufficient statistics.

The instrument response function and the thermal broadening were determined by non-dispersive and dispersive third order measurements of the 944.181 keV transition of $^{158}$Gd. The transition is very intense, depopulating a rather long lived state ($\tau > 5 \mbox{ps}$).  The results of this measurements are illustrated in Figure \ref{f3}. The instrument response function shows a deviation from the %theoretically 
  dynamical diffraction theory calculations, which mainly is due to vibrations of the crystals. This vibration amplitude is fitted and used in the further evaluation of dispersive scans. The extracted average thermal velocity was ${v_T=393(37)\mbox{m/s}}$, which is rather high compared to earlier experiments. This can be explained by the rather large sample mass of 9 grams causing a higher self-heating of the samples in the reactor.

\begin{figure}[ht!]
\includegraphics[width=\columnwidth]{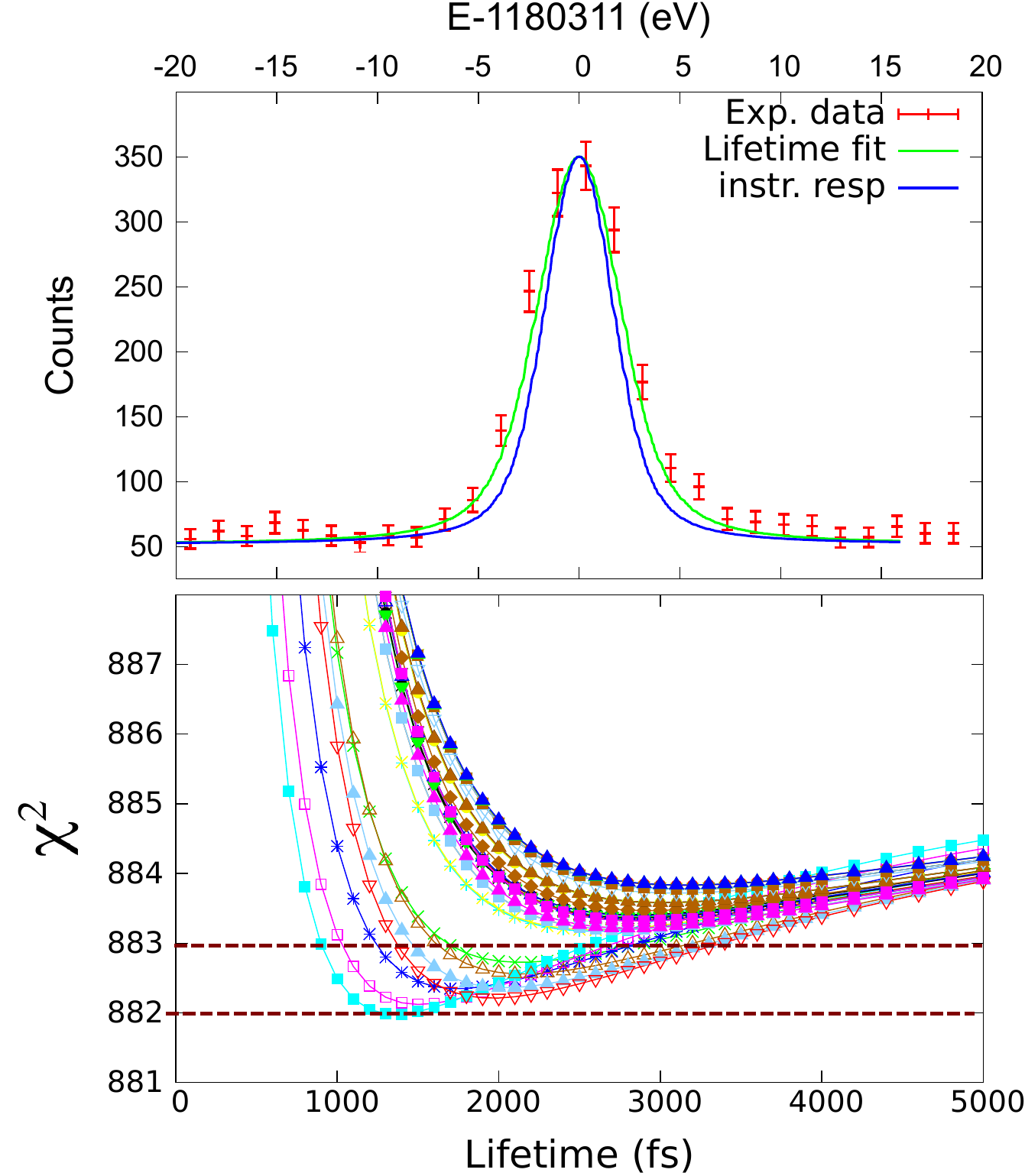}
\caption{\label{f4} The upper part shows the Doppler broadened line-shape of the 1180 keV transition depopulating the 4$^-$ state. The lower part shows the $\chi^2(\tau_{LOI})$ curves obtained for different $E_s, \tau_s$ combinations.  The horizontal dashed lines indicate the values of $\chi^2(\tau_{LOI})_{min}$ and $\chi^2(\tau_{LOI})_{min}+1$ allowing to find the most probable lifetime and to extract the error bar, respectively.}
\end{figure}

The measurement of the nuclear state lifetimes was performed by means of the measurement of supplementary Doppler broadening in dispersive third orders scans. Each gamma cascade,  populating the LOI, contributes to a recoil velocity distribution, which needs to be known for fitting the Doppler broadened lineshapes. The recoil velocity distribution is calculated using a Monte Carlo routine, following all possible feeding paths. It adds for each simulated gamma transition a recoil vector while also taking  into account  possible slowing down between consecutive recoil events. Only a small fraction (about 25$\%$) of the feeding of the 4$^-$ is known from the literature \cite{ensdf}. To our knowledge the published  (n,$\gamma$) level scheme has so far never been verified via coincidence measurements. Therefore it was verified with the results of the EXILL setup, which allowed at the present status of data evaluation a substantial correction/adding to the feeding scenario up to about 56$\%$. The remaining unknown part of the feeding was substituted by a virtual two step cascade. For this purpose  one assumes that the level of interest $E_{LOI}$ is connected with the capture state $E_{CAP}$ via a two step cascade over an intermediate level of energy $E_s$ with lifetime $\tau_s$. The values of these parameters are left free to minimise a global $\chi^2$. The range of variation for the energy was chosen to be  $E_{LOI}+200 \mbox{keV} < E_s < E_{CAP}-500 \mbox{keV}$, while the lifetime was allowed to vary between 5 to 500 fs. The Monte Carlo routine is applied to the combination of all known feeding cascades and the virtual  two step cascade. This  generates  a recoil velocity distribution, which is the basis of extracting a lifetime $\tau_{LOI}$ from the Doppler broadening data. For each parameter set $(E_s,\tau_s)$, a $\chi^2(\tau_{LOI})$ curve is generated. All simulated $(E_s,\tau_s)$  combinations yield a manifold of $\chi^2(\tau_{LOI})$ curves, which is used to extract the most probable lifetime and also to assign an error. The result for the 1180 keV transition is shown in the upper part of Figure \ref{f4}. The extracted lifetime is $\tau=1.4 \left(^{1.9}_{0.5} \right)$ ps. The same evaluation procedure was applied for the evaluation of the 1231 keV transition to extract the lifetime for the 2$^{-}$ state to be $2.9\left( ^{3.5}_{1.3} \right)$ps.

\section{Discussion}

A conversion of the obtained lifetime values into reduced branching ratios,  for the non-stretched E1-transitions, yields the values
\begin{eqnarray} 
 \mbox{B(E1)}[2^- \rightarrow 2^+] &=& 0.6\left(^{0.5}_{1.2} \right) \mbox{W.u.} \nonumber \\ 
 \mbox{B(E2)}[4^- \rightarrow 4^+] &=& 1.4\left(^{1.8}_{0.7} \right) \mbox{W.u., } \nonumber
\end{eqnarray}
whereas for the stretched E2-transition,
\begin{eqnarray}
\mbox{B(E2)}[4^- \rightarrow 2^-] &=& 1.0 \times 10^3 \left(^{1.3}_{0.6} \right)\mbox{W.u.} \nonumber
\end{eqnarray}

The deduced quadrupole moment of the even-spin negative-parity band is $Q_0=13.1\left( ^3 _{4.5} \right)$ b. 

The most surprising result here is certainly the  extraordinarily large B(E2)$[4^- \rightarrow 2^-]$,  which lies half way between the the B(E2) values obtained for normal deformed and super deformed structures, sign of high collectivity, for which at present we have no clear interpretation. It is worth mentioning that the present values of the lifetimes and the resulting branchings are still preliminary, since the evaluation of the EXILL data is not yet finished. However, since the EXILL data impact only the amount of known feeding the further evaluation will most likely only affect the error bars  and not shift the numbers of the extracted values. In this sense, it seems to be already clear that the difference of quadrupole moment of the odd and even spin negative parity bands indicates that these both bands should not be considered to be signature partners. Since the odd spin band is showing a quadrupole moment comparable with the ground state band ($Q_0=6.82$b) \cite{jen2010} it is questionable, whether the even spin band can be associated with   the pear-shape octupole vibration as suggested by some other authors \cite{kon81, cot96, sug11}.

\end{document}